\documentstyle [preprint,aps]{revtex}
\begin{document}
\preprint{gr-qc/9411032}

\title{STELLAR EQUILIBRIUM IN $2+1$ DIMENSIONS}
\addtocounter{footnote}{1}
\author {Norman Cruz$^{1,2}$\thanks{Electronic address: ncruz@lauca.usach.cl},
 Jorge Zanelli$^{1}$\thanks{Electronic address: jz@cecs.cl} }
\address{Centro de Estudios Cient\'{\i}ficos de Santiago, Casilla
16443, Santiago 9, Chile\\
 $^1$ Departamento de F\'{\i}sica, Facultad de Ciencias,
Universidad de Chile,\\
  Casilla 653, Santiago, Chile.\\
 $^2$ Departamento de F\'{\i}sica, Facultad de Ciencia,
Universidad de Santiago de Chile,\\
 Casilla 307, Santiago, Chile.}
\date{November 6, 1994}
\maketitle

\begin{abstract}
The hydrostatic equilibrium of a $2+1$ dimensional perfect
fluid star in asymptotically anti-de Sitter space is discussed.
The interior geometry matches the exterior $2+1$ black-hole solution.
An upper mass limit is found, analogous to Buchdahl's theorem in 3+1,
and the possibility of collapse is discussed. The case of a uniform matter
density is solved exactly and a new interior solution is presented.
\end{abstract}

In $3+1$ dimensions there is an upper bound on the ratio between
the mass and the radius for any general relativistic static fluid sphere:
$M/R \leq 4/9$.  Violation of this constraint would imply an
infinite central pressure, which signals a hydrostatic instability
\cite{buschdahl11}. This result shows that a sufficiently
massive star may become unstable and collapse, and the critical
mass ($M_c$) of the star is less than the Schwarzschild limit
(the condition for the spacetime signature to be preserved),
$M_c =4R/9 < R/2$. The inclusion of a cosmological constant
($\Lambda$) does not change the possibility of collapse
significantly \cite{hiscock22}.

On the other hand, a nonrelativistic star with uniform energy
density may have any size. The central pressure grows linearly
with the radius, there is no upper bound on $M/R$ and
therefore no collapse.

In $2+1$ dimensions and $\Lambda =0$, the radius of a static
fluid sphere depends solely on the mass density and not on the
central pressure \cite{giddings33}. This fact is a consequence
of the lack of gravitational attraction in that case. The
result found in \cite{giddings33} rules out the possibility of
gravitational collapse for $\Lambda = 0$.

Other results show that all structures in hydrostatic
equilibrium that obey a polytrope equation of state have a
gravitational mass equal to a constant (``1/2'') and produce
spacetimes of finite spatial volume with no exterior geometry
\cite{frankel44}.

These results are consistent with the absence
of an analog of the Schwarzschild geometry in $2+1$ dimensions
without cosmological constant.

Point mass solutions with nonzero cosmological constant present
horizons and the radius of curvature $l=(-\Lambda)^{-
{1\over2}}$ provides the necessary length scale. If $\Lambda
\geq 0$ a cosmological type horizon exist with a naked
singularity.

A black-hole solution has been recently found in $2+1$
dimensions with negative cosmological constant. It exhibits
thermodynamic properties similar to those for 3+1 black holes.
\cite{banados55}. The existence of this black hole suggests
the possibility that a fluid distribution in $2+1$ dimensions
could collapse, as it happens with four dimensional stars of a
mass greater than the upper limit for cold matter in hydrostatic
equilibrium at the end of their thermonuclear phase. The possibility
of collapse into a black hole final state has been shown to occur for a
spherically symmetric distribution of pressureles fluid (dust)
\cite{mann66}.

Here we explore the stability of a spherical
distribution of matter described by a generic equation of
state relating the pressure ($p$) and energy density ($\rho$):
$p=p(\rho)$, $\rho \leq \rho_o$. This ($2+1$)- dimensional star is
assumed in hydrostatic equilibrium and the exterior geometry is
asymptotically anti-de Sitter. An upper mass limit is found
under the following additional assumptions. i) the energy
density is positive semi definite: $\rho \geq 0$, and, ii)
matter is microscopically stable: ${dp\over d\rho} \geq 0$.

Consider Einstein's equations with cosmological constant for a
perfect fluid with energy density $\rho$ and pressure
$p$.\footnote{ We take the gravitational action normalized as
${1 \over {2k}}\int{\sqrt{-g} Rd^3x}$, and $G = c =\hbar = 1$.}

\begin{equation}
\frac{1}{k} (R_{\mu \nu} - {1 \over 2} g_{\mu \nu} R +
\Lambda g_{\mu \nu}) =
(\rho + p) u_\nu u_\mu - pg_{\mu \nu},
\label{einstein}
\end{equation}
where $k$ is the gravitational constant with dimensions of $[mass]^{-1}$.

Taking the spherically symmetric, static $2+1$ metric $ds^2 =
e^{2 \nu (r)}dt^2 - e^{2 \lambda(r)}dr^2 - r^2 d \phi^2$,
Equations (\ref{einstein}) become

\begin{equation}
e^{-{2 \lambda(r)}} \lambda' = k (\rho + \Lambda) r
\label{e-lamda}
\end{equation}
and

\begin{equation}
e^{-{2 \lambda(r)}} \nu' = k (p - \Lambda) r
\label{e-nu}
\end{equation}
Equation (\ref{e-lamda}) is  easily integrated into

\begin{equation}
e^{-2 \lambda} = C - \frac{k}{\pi} \mu(r;\Lambda),
\label{sol-lam}
\end{equation}
where

\begin{eqnarray}
\mu(r)  =\int_0^r 2 \pi (\rho(r') + \Lambda)r'dr' \nonumber \\
= m(r) + \pi \Lambda r^2
\label{mu}
\end{eqnarray}

and the integration constant $C$ must be such that the origin be
part of the spacetime --i.e., $C> \frac{k}{\pi} \mu(r;\Lambda)$,
for $0\leq r \leq R$, where $R$ is the radius of the star--, and
it is fixed by matching with the exterior black hole metric
given in \cite{banados55}

The above equations combined with the conservation of stress
energy yields the corresponding Oppenheimer -Volkov equation in
$2+1$ dimensions

\begin{equation}
{dp \over dr} = {-k (p + \rho) (p - \Lambda) r \over C -
\frac{k}{\pi} \mu(r)}
\label{O-V}
\end{equation}
which can be integrated, given the usual boundary condition for
the central pressure $p(r = 0) = p_c$. The radius of the star
is defined by the condition $p(r = R) = 0$, and the mass of the
fluid is

\begin{eqnarray}
M=: \mu(r = R;\Lambda = 0) = 2\pi \int _0^R \rho(r) rdr \nonumber \\
= \mu(R) - \pi \Lambda r^2.
\label{mass}
\end{eqnarray}

The denominator of equation (\ref{O-V}) is $g_{11}$ and hence it
must be positive at surface of the star. This sets an upper bound
on $M$ given by

\begin{equation}
M \leq  \pi \left [{C \over k} - \Lambda R^2 \right].
\label{masslim}
\end{equation}

This condition derives from the requirement that the the metric signature be
maintained throughout spacetime, and is therefore the analog of the condition
$M < R/2$ in the Schawarzchild case.

The behaviour of perfect fluids in hydrostatic equilibrium in
$2+1$ dimensions can be described using the following general
results, which are valid irrespective of the equation of state
considered.

{\it LEMMA 1: A perfect fluid in hydrostatic equilibrium is only
possible for $\Lambda \leq 0$}.

Proof: Evaluating eq.(\ref{O-V}) at the surface of the sphere
where $p=0$ we obtain

\begin{equation}
\left. \frac{dp}{dr}\right|_{r=R} = {k \Lambda \rho R \over C- \frac{k}{\pi}
\mu(R)}.
\label{lem1}
\end{equation}

The denominator is positive at the surface of the fluid sphere.
But since $p(r<R) >0$, the pressure gradient must be non
positive at the surface, we conclude that $\Lambda \leq 0$.$\Box$

Since we are interested in fluids near hydrostatic equilibrium,
we consider only the case with a negative cosmological constant.
For simplicity we shall write $\Lambda$ as $- 1 \over l^2$.

{\it LEMMA 2: The central pressure of a fluid in hydrostatic
equilibrium is bounded from below by}

\begin{equation}
p_c > {k \over 2 \pi l^2 C} M
\label{lem21}
\end{equation}

Proof: From the eq.(\ref{O-V}) and the expresion for $\rho$ in terms of
the mass $m(r)$ given by the equation

\begin{equation}
{dm \over dr} = 2 \pi \rho(r)
\label{lem22}
\end{equation}

we obtain

\begin{equation}
{d \over dr} \left[p \left(m(r) + {\pi C \over k} \right) +
{m(r) \over 2l^2} + {r^2p \over l^2} \right] = -\pi p \left(p +
l^2 \right)r + 2 m(r) {dp \over dr} + {2rp \over l^2}.
\label{lem23}
\end{equation}

The right hand side of eq. (\ref{lem23}) is always negative. Thus,
the function given by
\begin{equation}
\Re = p(r) \left(m(r) + {\pi \over k} \right) + {m(r) \over 2l^2} + {r^2p
\over l^2}
\label{funR}
\end{equation}
is monotonically decreasing with $r$, and $\Re (r = 0) > \Re (r =
R)$. This leads to the lower bound given by (\ref{lem21}).$\Box$\\

Equation (\ref{lem21}) expresses the simple fact that, for negative $\Lambda$,
as the mass grows so does the central pressure needed to support it in
gravitational equilibrium.

Buchdahl has shown that in $3+1$ dimensions, the equations of
relativistic stellar structure for cold matter leads to the
existence upper mass limit which is above the naive limit
obtained if the radius of the star is Schwarzschild's, $M/R=1/2$
\cite{buschdahl11},

\begin{equation}
{M \over R} \leq {4 \over 9}.
\label{MR}
\end{equation}

The bound (\ref{MR}) is valid irrespective of the equation of
state provided that: i) matter is described by a one parameter
equation of state relating $p$ and $\rho$: $p=p(\rho)$; ii) the
density is positive and monotonically decreasing (${d\rho \over
dr} < 0$); iii) matter is microscopically stable (${dp\over
d\rho} \geq 0$).

Under the same assumptions, the following bound on $\mu(r)$
is found in $2+1$ dimensions (see Appendix)

\begin{equation}
\mu(r) \leq {\pi C \over 2k} \left [1 - {2k \over C}(p -
\Lambda)r^2 + \sqrt {{4k \over C}(p -\Lambda)r^2 + 1} \right].
\label{buchlim}
\end{equation}

As we see from LEMMA 1, hydrostatic equilibrium is only possible
when $\Lambda \leq 0$. For $\Lambda = 0$ we require $\mu(R) =
M \leq \pi/k$ in order to preserve the metric signature and in
that case, (\ref{buchlim}) does not give a better bound. Setting
$- \Lambda = 1/l^2$ and evaluating (\ref{buchlim}) at the surface of the
fluid, one finds that te mass of the fluid satisfies

\begin{equation}
M = \mu(R) + \frac{\pi R^2}{l^2} \leq {\pi C \over 2k} \left [1
+ \sqrt {{4k R^2 \over C l^2} + 1} \right],
\label{buchm}
\end{equation}
which is above the naive limit (\ref{masslim}), and approaches that
value as $R\rightarrow 0$.

The general relativistic equation of hydrostatic equilibrium (\ref{O-V})
can be integrated exactly when the fluid density is uniform.
Integrating Einstein's equations for $\rho_o = const$ yields

\begin{equation}
p(r) = \frac{\rho_o \left [(C-{\pi \over k} \mu(r))^{1/2} - (C-{\pi \over k}
\mu(R))^{1/2} \right ]}{l^2 \rho_o (C-{\pi \over k} \mu(R))^{1/2} - (C-{\pi
\over k} \mu(r))^{1/2}}
\label{ro-cte}
\end{equation}

The central pressure required for equilibrium of a uniform density fluid is

\begin{equation}
p_c = \frac{\rho_o \left [C - (C-{\pi \over k} \mu(R))^{1/2} \right]}{l^2
\rho_o \left[C-{\pi \over k} \mu(R) \right]^{1/2} - C}
\label{pc}
\end{equation}
$p_c$ becomes infinite when

\begin{equation}
\mu(R) = {\pi \over k} \left [ C - \left ( {C \over l^2 \rho_o}
\right )^2 \right ]
\label{ro-pc}
\end{equation}
or, in terms of the mass $M$

\begin{equation}
M = {\pi \over k} \left [ C - \left ( {C \over l^2 \rho_o} \right
)^2 + {kR^2 \over l^2} \right ]
\label{m-pc}
\end{equation}

The exterior solution is obtained taking $\rho = p = 0$, which
is to be compared with the black-hole metric \cite{banados55},

\begin{equation}
ds^2 = - (- M_o + {r^2 \over l^2}) dt^2 + (- M_o + {r^2 \over l^2})^{-2}
dr^2 + r^2 d \phi^2
\label{2+1}
\end{equation}
with $-\infty < t < \infty\;,\; 0 < r < \infty$, and $0 \leq \phi \leq 2\pi$.
This metric was obtained taking $k = \pi$. We will use this value in the follo
wing equations. The parameter $M_o$ is the conserved charge associated with
asymptotic invariance under time displacements (mass). This charge is given by
a flux integral through a large circle at spacelike infinity.

The continuity of the metric at the surface of the star determines
the relation between $M$ and the conserved charge $M_o$,
associated to the asymptotic Killing vector of the black hole
geometry,

\begin{equation}
M_o = M - C
\label{mcero}
\end{equation}

This relation is consistent with the vacuum state considered in
\cite{banados55}, obtained by making the black hole disappear,
that is, by letting the horizon size go to zero.

The interior metric, $e^{2 \nu}$, is obtained from equations
(\ref{e-nu}) and (\ref{sol-lam}) when the expresion for $p(r)$, given
in eq(\ref{ro-cte}), is considered

\begin{equation}
e^{2 \nu} =  \left [ l^2 \rho_o (1-{\pi C \over k} \mu(R))^{1/2} - (1-{\pi C
\over k} \mu(r))^{1/2} \over {l^2 \rho_o - 1} \right ]^2 \left [{-M_o + {r^2
\over l^2}} \over (1-{\pi C \over k} \mu(r))^{1/2}\right ].
\label{inmetric}
\end{equation}
\vspace{1cm}

{\bf Discussion and Conclusions}

We have found that the presence of a negative cosmological
constant opens up the possibility for a fluid sphere in $2+1$
dimensions to collapse. This gravitational attraction, however,
cannot be attributed to the presence of a dynamical field, as is
well known from the fact that in three dimensions the Weyl
tensor vanishes and at each point the curvature tensor is completely
determined by the energy-momentum content at that point.

One can observe the difference between three and four dimensions
comparing (\ref{O-V}) with the corresponding equation for
hydrostatic equilibrium for $D=4$,

\begin{equation}
{dp \over dr} = - (\rho + p) {m(r) + 4 \pi r^3(p -\Lambda) \over
r^2 (1- 2m(r)/r)}.
\label{O-V4}
\end{equation}

This expression depends on the total mass inside the radius $r$,
whereas the pressure gradient in eq.(\ref{O-V}) depends on the
pressure and density at the same point.

It is remarkable that eqs.(\ref{O-V4}) and (\ref{O-V}) have the
same form for constant $\rho_o$. In both cases this implies the
existence of an upper mass limit and the inevitability of
collapse beyond that limit.

If we consider the black hole metric as the exterior of a
$2+1$ dimensional star, an upper bound on the mass is given by
the equivalent of Buchdahl's theorem \cite{buschdahl11} (see
appendix)

\begin{equation}
{M_o} \leq M_c =:{1 \over 2} \left [ (C^2 + {4 \pi C R^2 \over
l^2})^{1/2} - C \right ]
\label{lim-mo}
\end{equation}

This upper bound is more restrictive than the one obtained by
requiring the metric coefficient $g_{oo}$ to be always negative
for $r \geq R$.

If the cosmological constant were not included, all rotationally
invariant stable structures have the same mass \cite{cornish88}.

For a star of uniform density the critical mass, $M_c$, becomes

\begin{equation}
M_c = {\pi R^2 \over l^2} - \left ( {C \over l^2 \rho_o} \right )^2.
\label{mo-rocte}
\end{equation}

In particular, equation (\ref{mo-rocte}) implies that there is
a minimum radius for a $2+1$ dimensional star with uniform density,

\begin{equation}
R \geq {C \over \pi^{1/2}} \left({1 \over l \rho_o }\right).
\label{Rmin}
\end{equation}

It is not a completely obvious result that the $2+1$ dimensional
black hole solution could be matched to a perfect fluid star.
Hydrostatic equilibrium imposes limits on the mass and radius
beyond which the star would undergo collapse. Without this
result, the study of a gravitationally collapsing disk of dust
can only show that {\em if} the dust is falling, a black hole
will be formed; but it doesn't prove that matter left to its own
will necessarily collapse.

The study of the maximal analytic extension of the black hole
solution shows, as in the four dimensional case, non-physical
regions. If we assume that these extensions describe complete
gravitational collapse, collapsing matter would cover up these
non-physical parts. Thus, one can argue that the gravitational
collapse of a spherical body always produces a black hole in
$2+1$ dimensions.

\acknowledgments

This work was partially supported by Grants 193.0910/93,
194.0203/94 and 2930007/93 from FONDECYT (Chile), grant
04-9331CM from DICYT (Universidad de Santiago), by a European
Communities research contract, by institutional support to the
Centro de Estudios Cient!ficos de Santiago provided by SAREC
(Sweden), and a group of chilean private companies (COPEC,
ENERSIS, CGEI). This research was also sponsored by CAP IBM and
XEROX-Chile.

\appendix{Appendix}
\renewcommand{\theequation}{A.\arabic{equation}}
\setcounter{equation}0

In this appendix we establish the upper bound (\ref{buchlim}).
{}From eqs.(\ref{e-nu}), (\ref {sol-lam}) and (\ref{O-V}) we
find

\begin{equation}
{d \over dr} \left [\frac{\nu'}{r} (\beta - 2\mu) \right] +
\left [(\beta - 2\mu ) \nu' + \mu' \right ] \frac{\nu'}{r} = 0
\label{A1}
\end{equation}
where $\beta = \frac{2 \pi C}{k}$. If we define the independent
variables

\begin{equation}
\xi(r) = : e^{\nu(r)}
\label{A2}
\end{equation}
and
\begin{equation}
\gamma = \int_o^r d r' r' (\beta-2\mu)^{-1/2},
\label{A3}
\end{equation}
equation (\ref{A1}) reads

\begin{equation}
{d^2 \xi \over d \gamma ^2} = 0,
\label{A4}
\end{equation}
which can be integrated as

\begin{equation}
\xi(r) = {d\xi \over d\gamma} \gamma + \xi(0).
\label{A5}
\end{equation}
Since both $\xi(0)$ and $\gamma$ are non-negative, we have from (\ref{A5})

\begin{equation}
{1 \over \xi} {d \xi \over d \gamma} \leq {1 \over \gamma}.
\label{A6}
\end{equation}
Thus, in terms of $r$ and $\nu (r)$, eq.(\ref{A6}) reads

\begin{equation}
{(\beta - 2 \mu(r))^{1/2} \over r} {d \nu \over dr} \leq \left [
\int^r_0 dr' r' (\beta- 2 \mu(r'))^{-1/2} \right ] ^{-1}.
\label{A6'}
\end{equation}

The fact that the matter density is monotonically decreasing
with $r$ (${d \rho \over dr} < 0$), implies that the average
density is also decreasing, namely

\begin{equation}
{d \over dr} \left (\mu \over r^2 \right ) < 0.
\end{equation}
\label{A7}
This in turn implies that
\begin{eqnarray}
\int^r_0 d r' r' (\beta - 2 \mu(r'))^{-1/2} \geq \int^r_0 dr' r'
\left (\beta- {2 \mu(r) \over r^2} r'^2 \right )^{-1/2}
\nonumber \\
= {r^2 \over 2 \mu(r)} \left [ \beta^{1/2}- (\beta-2
\mu(r))^{1/2} \right ],
\label{A8}
\end{eqnarray}
and (\ref{A6'}) becomes
\begin{equation}
(1-x)^{1/2} \nu' \leq {x \over r}[1 -\sqrt{1-x}]^{-1},
\label{A9}
\end{equation}

where $x =:{2\mu(r) \over \beta} = {k \over \pi C}\mu(r)$. Using
eqs.(3) and (4), the left hand side of eq.(\ref{A9}) can be
written as $ {k \over C}r (p+ {1\over l^2})(1-x)^{-1/2}$, and
therefore

\begin{equation}
\left[ \frac{k}{C}(p + {1 \over l^2}) +x-1 \right]^2 \leq 1-x,
\label{A10}
\end{equation}
which, after a little algebra gives the upper bound on $\mu(r)$
(\ref{buchlim}).

\end{document}